# *Insight*-HXMT Observations of Swift J0243.6+6124 during Its 2017–2018 Outburst

Yue Zhang[1,2], MingYu Ge[1], LiMing Song[1], ShuangNan Zhang[1,2,3], JinLu Qu[1], Shu Zhang[1], Victor Doroshenko[4,5], Lian Tao[1],
Long Ji[4], Can Güngör[1], Andrea Santangelo[4], ChangSheng Shi[6], Zhi Chang[1], Gang Chen[1], Li Chen[7], TianXiang Chen[1],
Yong Chen[1], YiBao Chen[8], Wei Cui[1,8], WeiWei Cui[2], JingKang Deng[8], YongWei Dong[1], YuanYuan Du[1], MinXue Fu[8],
GuanHua Gao[1,2], He Gao[1,2], Min Gao[1], YuDong Gu[1], Ju Guan[1], ChengCheng Guo[1,2], DaWei Han[1], Wei Hu[1], Yue Huang[1],
Jia Huo[1], ShuMei Jia[1], LuHua Jiang[1], WeiChun Jiang[1], Jing Jin[1], YongJie Jin[9], Bing Li[1], ChengKui Li[1], Gang Li[1], MaoShun Li[1],
Wei Li[1], Xian Li[1], XiaoBo Li[1], XuFang Li[1], YanGuo Li[1], ZiJian Li[1,2], ZhengWei Li[1], XiaoHua Liang[1], JinYuan Liao[1],
CongZhan Liu[1], GuoQing Liu[8], HongWei Liu[1], ShaoZhen Liu[1], XiaoJing Liu[1], Yuan Liu[3], YiNong Liu[9], Bo Lu[1], XueFeng Lu[1],
Tao Luo[1], Xiang Ma[1], Bin Meng[1], Yi Nang[1,2], JianYin Nie[1], Ge Ou[1], Na Sai[1,2], Liang Sun[1], Ying Tan[1], WenHui Tao[1],
YouLi Tuo[1,2], GuoFeng Wang[1], HuanYu Wang[1], Juan Wang[1], WenShuai Wang[1], YuSa Wang[1], XiangYang Wen[1], BoBing Wu[1],
Mei Wu[1], GuangCheng Xiao[1,2], ShaoLin Xiong[1], He Xu[1], YuPeng Xu[1], LinLi Yan[10], JiaWei Yang[1], Sheng Yang[1], YanJi Yang[1],
AiMei Zhang[1], ChunLei Zhang[1], ChengMo Zhang[1], Fan Zhang[1], HongMei Zhang[1], Juan Zhang[1], Tong Zhang[1], Wei Zhang[1,2],
WanChang Zhang[1], WenZhao Zhang[7], Yi Zhang[1], YiFei Zhang[1], YongJie Zhang[1], Zhao Zhang[8], ZiLiang Zhang[1], HaiSheng Zhao[1],
JianLing Zhao[1], XiaoFan Zhao[1,2], ShiJie Zheng[1], Yue Zhu[1], YuXuan Zhu[1], and ChangLin Zou[1]
(Insight-HXMT collaboration)

[1] Key Laboratory of Particle Astrophysics, Institute of High Energy Physics, Chinese Academy of Sciences, 19B Yuquan Road, Beijing 100049, People's Republic of China; zhangyue@ihep.ac.cn, songlm@ihep.ac.cn
[2] University of Chinese Academy of Sciences, 19A Yuquan Road, Beijing 100049, People's Republic of China
[3] National Astronomical Observatories, Chinese Academy of Sciences, Beijing 100012, People's Republic of China
[4] Institut für Astronomie und Astrophysik, Kepler Center for Astro and Particle Physics, Universität Tübingen, Sand 1, D-72076 Tübingen, Germany
[5] Space Research Institute of the Russian Academy of Sciences, Profsoyuznaya Str. 84/32, Moscow 117997, Russia
[6] College of Material Science and Chemical Engineering, Hainan University, Hainan 570228, People's Republic of China
[7] Department of Astronomy, Beijing Normal University, Beijing 100088, People's Republic of China
[8] Department of Physics, Tsinghua University, Beijing 100084, People's Republic of China
[9] Department of Engineering Physics, Tsinghua University, Beijing 100084, People's Republic of China
[10] School of Mathematics and Physics, Anhui JianZhu University, Hefei 230601, People's Republic of China

## Abstract

The recently discovered neutron star transient Swift J0243.6+6124 has been monitored by the *Hard X-ray Modulation Telescope*. Based on the obtained data, we investigate the broadband spectrum of the source throughout the outburst. We estimate the broadband flux of the source and search for possible cyclotron line in the broadband spectrum. However, no evidence of line-like features is found up to 150 keV. In the absence of any cyclotron line in its energy spectrum, we estimate the magnetic field of the source based on the observed spin evolution of the neutron star by applying two accretion torque models. In both cases, we get consistent results with $B \sim 10^{13}$ G, $D \sim 6$ kpc and peak luminosity of $>10^{39}$ erg s$^{-1}$, which makes the source the first Galactic ultraluminous X-ray source hosting a neutron star.

*Key words:* accretion, accretion disks – pulsars: individual (Swift J0243.6+6124) – stars: distances – stars: magnetic field – X-rays: binaries

## 1. Introduction

Neutron star X-ray binaries are binary systems consisting of a magnetized neutron star accreting matter supplied by a nondegenerate stellar companion. The observed X-ray emission is powered by accretion of captured material funneled by the strong magnetic field onto the magnetic poles of the neutron star. Meanwhile, the neutron star also accretes the angular momentum carried by the accretion flow. Variations of the spin-up rate are thus correlated with the mass accretion rate (see Ghosh & Lamb 1979; Wang 1995; Kluźniak & Rappaport 2007; Shi et al. 2015, and references therein).

The transient X-ray source Swift J0243.6+6124 was discovered on 2017 October 3 by *Swift*/BAT telescope, and it was suggested that the compact object is a neutron star (Kennea et al. 2017). X-ray pulsations were also detected with a period of ∼9.86 s (Bahramian et al. 2017; Jenke & Wilson-Hodge 2017; Kennea et al. 2017) modulated by the motion in an eccentric orbit ($e \sim 0.1$) with a period of ∼28 days (Ge et al. 2017; Doroshenko et al. 2018). Based on the observation from the 1.3 m telescope of the Skinakas Observatory, the optical counterpart of the source has been identified as a Be star (Kouroubatzakis et al. 2017), thus confirming the system as a Be/X-ray binary (BeXRB). The distance to the companion star was estimated at 2.5 kpc using spectro-photometry of the Be counterpart (Bikmaev et al. 2017). On the other hand, Doroshenko et al. (2018) showed that the minimum distance must be 5 kpc to explain the observed spin-up rate. Analysis of the spin evolution provided also estimates the magnetic field at $B \sim 10^{13}$ G. Subsequently, the distance estimate was confirmed as $7.3^{+1.6}_{-1.2}$ kpc using the measured parallax given by the *Gaia* Observatory (van den Eijnden et al. 2018).

The observed high flux implies for such distance that the peak luminosity is up to $\sim 3 \times 10^{39}$ erg s$^{-1}$, which leads to the classification of this source as the first Galactic ultraluminous

X-ray (ULX) source (Tsygankov et al. 2018). As discussed by van den Eijnden et al. (2018), even assuming the lower limit for the distance, the peak of $1.1 \times 10^{39}$ erg s$^{-1}$ implies that the Eddington limit for the neutron star was exceeded during the outburst. Moreover, optically thick outflows found in *NuSTAR* observations also confirm that the source is a super-Eddington accretion system (Tao et al. 2019).

The newly launched X-ray astronomical satellite *Hard X-ray Modulation Telescope* (*Insight-HXMT*)[11] conducted the monitoring campaigning of this source starting on 2017 October 7. It is the first X-ray astronomical satellite of China, based on the Direct Demodulation Method (Li & Wu 1993, 1994) and was launched on 2017 June 15. There are three main payloads carried by *Insight-HXMT* (Zhang et al. 2014): the High Energy X-ray telescope (HE) with a total detection area of 5100 cm$^2$ in the energy range 20–250 keV, the Medium Energy X-ray telescope (ME) with a total detection area of 952 cm$^2$ in the energy range 5–30 keV, and the Low Energy X-ray telescope (LE) with a total detection area of 384 cm$^2$ in the energy range 1–15 keV. The recent progresses around this telescope can be found in Zhang et al. (2018), Li et al. (2018), Jia et al. (2018), Chen et al. (2018), Huang et al. (2018), and Tao et al. (2019). For the current study, its large effective area in the broadband energy range of 1–250 keV, and flexible scheduling of the observations are of particular importance.

In this paper, we report the results of the analysis of the *Insight-HXMT* data. The observation information and data analysis are described in Section 2. In Section 3, we present the estimation of the magnetic field using two accretion torque models. Finally, we give a discussion and summarize our study in Section 4.

## 2. Observations and Data Analysis

Swift J0243.6+6124 was observed 98 times by *Insight-HXMT* in pointed observation mode starting on 2017 October 7; observations with a typical duration of 10 ks were scheduled every 1–2 days between MJD 58,033 and MJD 58,170. In total 98 individual pointing observations with a total net exposure time of ~1205 ks are obtained.

### 2.1. Data Reduction

The data are reduced following standard procedures using the *Insight-HXMT* data analysis software package HXMTDAS v2.01. The details of data analysis procedures are reported in the HXMTDAS documentation.[12] However, the main steps can be summarized as follows:

1. To generate the calibrated events from the raw events according to the Calibration Database (CALDB) files using the HXMTDAS tasks of hepical, mepical, and lepical for data of HE, ME, and LE instruments, respectively.
2. Using a given screening criterion to generate the Good Time Intervals (GTIs) file for each of the detectors using hegtigen, megtigen, and legtigen tasks.
3. Extracting events from the calibrated events according to GTIs file using hescreen, mescreen, and lescreen tasks.
4. Extracting source spectra from screened events with hespecgen, mespecgen, and lespecgen tasks.
5. Calculating background spectra with screened events by hebkgmap, mebkgmap, and lebkgmap tasks.
6. Creating the response matrix and ancillary response file with herspgen, merspgen, and lerspgen.

The screening criteria parameters mainly include the Earth elevation angle (ELV), the cutoff rigidity (COR), the offset angle from the pointing direction (ANG_DIST) and the South Atlantic Anomaly Flag (SAA_FLAG). In our data analysis procedure, the extracted screened events are limited to the COR more than eight for each detector to eliminate charged particle contribution. Some events, taken during satellite slews and passages through the South Atlantic Anomaly, were filtered out. Additionally, we also exclude the events with low ELV to limit the background level; the critical value to constrain the events are chosen as 10° for HE and ME, and as 15° for LE. The LE instrument parameter of Bright Earth Angle is also set as more than 40° to limit the background.

The arrival times of all the screened events are referred to the solar system barycenter to estimate accurate ephemeris of the observation, because of the motion of the satellite and the Earth. This step is done by using the HXMTDAS tool hxbary which uses the orbital information to reconstruct the arriving time and DE-405 ephemeris for Earth motion. We also assume the position of the source reported by Kennea et al. (2017).

### 2.2. Spectral Analysis

The presented spectral analysis is highly preliminary as both the software and calibration of *Insight-HXMT* are still in active development. The pulse averaged spectral analysis is performed for all observations in the 2–150 keV range. The corresponding background spectra are estimated multiplying by the count rate of the blind field of view (FoV) detectors following the procedure in Section 2.1. The value of the multiplication factor is the ratio of the number of nonblind FoV detectors to that of the blind FoV detectors. This method is tested by the *Insight-HXMT* background team using blank sky observations. Besides the resulting spectrum, we also use the fact that the source pulsates and use the off-pulse spectrum (extracted from the screened events in the lowest intensity phase bin in the pulse profile) as an estimate of the background spectrum of the pulse-on spectrum.

The pulse averaged spectra and the pulse-on spectra are fitted using XSPEC package version 12.10.0 c (Arnaud 1996) with different models. For the latter case, the spectra are approximately reproduced by cons*TBabs*(cutoffpl+bbody) with a systematic error of 0.5%, that accounts for residual calibration uncertainties. Interstellar absorption is accounted for by the model TBabs with abundances from Wilms et al. (2000). On the other hand, with the addition of the gaussian profile model (i.e., the model is cons*TBabs*(cutoffpl+bbody+Gauss)) to describe the iron emission line, the pulse averaged spectra can be approximately reproduced assuming the same systematic error of 0.5% (Figure 1 and Table 1). The distribution of the best-fitting reduced $\chi^2$ is shown in Figure 2. It should be noted that the reduced $\chi^2$ is relatively larger in a few cases. In order to examine the accuracy of flux estimation in those observations, we used the values of estimated flux from the nearest well-fitted observations and interpolated them as a function of the count rate to calculate the derived flux. We found <4% difference between the values of estimated flux and derived flux. For instance, in the case of ObsID

---

[11] http://www.hxmt.org
[12] http://www.hxmt.org/index.php/enhome/analysis/199-hxmt-data-anslysis-software

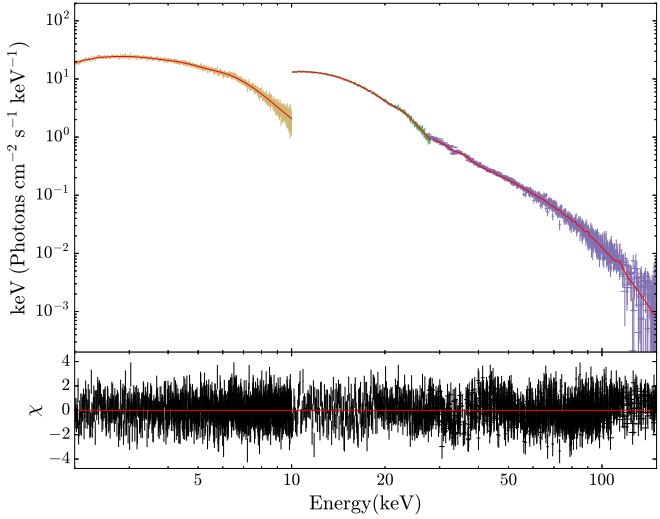

**Figure 1.** Fitting result of the pulse averaged energy spectrum of Swift J0243.6+6124 in the 2–150 keV range observed with the HE (purple dotted line), ME (green dotted line), and LE (yellow dotted line) instruments of *Insight-HXMT* on MJD 58147. The systematic error is fixed at 0.5% throughout the outburst. The spectrum is reproduced with the model `cons*TBabs*(cutoffpl+bbody+Gauss)`. The reduced $\chi^2$(dof) is 0.92(2686), the best-fitting spectral parameters of this observation is given in Table 1.

**Table 1**
Spectral Parameters of the Best-fit Model for the Observation Taken on MJD 58,147

| Component | Parameter | Value |
|---|---|---|
| TBabs | $N_{\rm H}$ ($10^{22}$ cm$^{-2}$) | $0.88 \pm 0.08$ |
| cutoffpl | Photon index | $1.25 \pm 0.02$ |
|  | $E_{\rm cut}$ (keV) | $28.35^{+0.48}_{-0.47}$ |
|  | norm | $0.73 \pm 0.03$ |
| bbody | $T_{\rm bb}$ (keV) | $3.36^{+0.05}_{-0.04}$ |
|  | norm | $0.031 \pm 0.002$ |
| gaussian | $E_{\rm g}$ (keV) | $6.98 \pm 0.15$ |
|  | $\sigma$ (keV) | $1.04^{+0.22}_{-0.23}$ |
|  | norm | $0.151^{+0.005}_{-0.004}$ |

**Note.** The reduced $\chi^2$(dof) is 0.92(2686).

O011457701036 with a maximum reduced $\chi^2$ (1.57), we used its nearest observation O011457701035 ($\chi^2 = 1.23$) to calculate the derived flux of ObsID O011457701036 and found ∼1% difference from its estimated flux value. What shall be insisted is that the spectral fitting is only applied to flux estimation in this work, and the detailed spectra components are not discussed here. (Detailed spectral studies using *NuSTAR* observations can be found in Tao et al. 2019).

In Figure 3, we present the bolometric light curve of Swift J0243.6+6124 derived for both spectral models. The total flux increased from the beginning, reached the maximum value at MJD 58065, and then began to decrease smoothly. The total flux thus changed by a factor of more than 100 from ∼$2.5 \times 10^{-9}$ to ∼$3.3 \times 10^{-7}$ erg cm$^{-2}$ s$^{-1}$ within the epoch covered by observations. At the same time, the pulsed flux evolution shows a similar behavior, which changed from ∼$1.6 \times 10^{-9}$ to ∼$2.1 \times 10^{-7}$ erg cm$^{-2}$ s$^{-1}$.

Given that energy calibration of the *Insight-HXMT* is still in progress, we use the ratio of the observed spectrum of the source to that of the Crab pulsar (the spectral ratio) to search for narrow features associated with a possible cyclotron line. The function `TBabs*(cutoffpl+bb)` is used to fit the spectral ratio (Figure 4). No such features are, however, found in any of the observations between 2 and 150 keV (the corresponding magnetic field for this energy range is from $1.7 \times 10^{11}$ G to $1.3 \times 10^{13}$ G). This result covers a broader energy range compared to the result of Jaisawal et al. (2017) from 3 to 79 keV with *NuSTAR*. Nondetection of the line in *Insight-HXMT* agrees with previous conclusions and suggests that either the line is not generated for $B < 1.3 \times 10^{13}$ G, or the magnetic field is stronger than that.

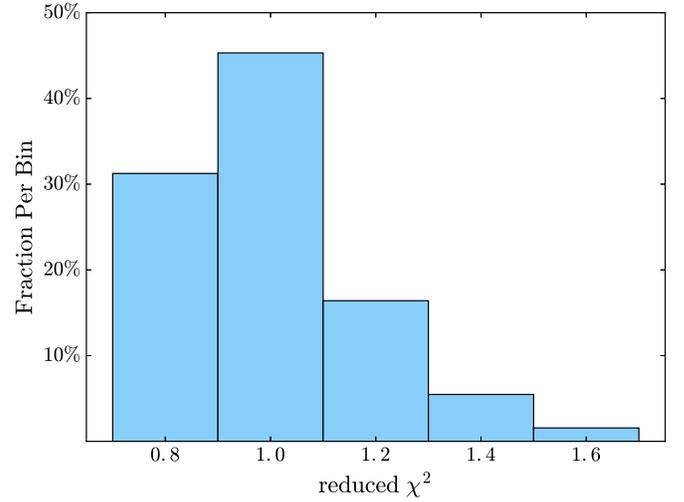

**Figure 2.** Distribution of the pulse averaged spectra fitting result (reduced $\chi^2$) in 2–150 keV range *Insight-HXMT* data obtained with XSPEC. Most of the observed spectra can be approximately well reproduced.

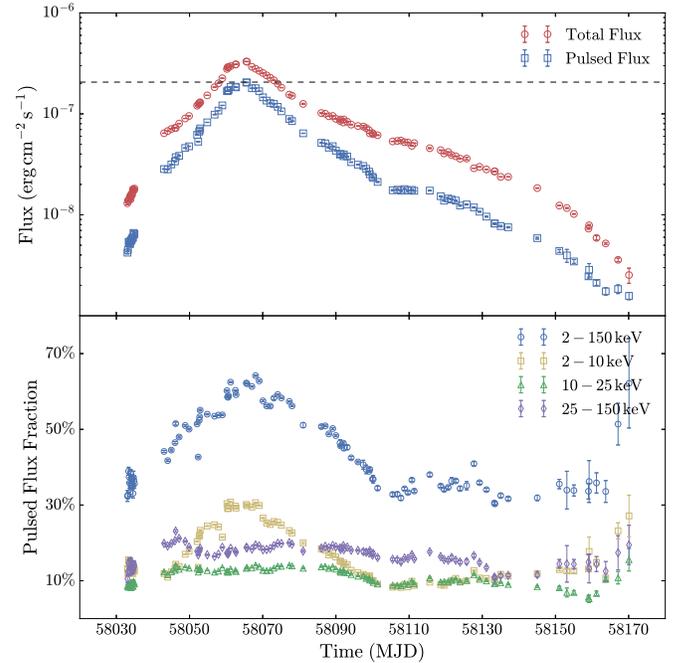

**Figure 3.** Top: the bolometric flux (denoted by red open circles) and the pulse-on flux (denoted by blue squares) are estimated by fitting *Insight-HXMT* spectra. The dotted line denotes the Eddington limit in flux using a distance of 5 kpc. Bottom: the pulsed flux fraction of different energy bands. The pulsed flux fraction of the whole energy range (2–150 keV) is denoted by blue open circles. Purple open diamonds represent the HE part of the pulsed flux fraction; green open triangles represent the ME part of the flux fraction; yellow open squares represent the LE part of the flux fraction.

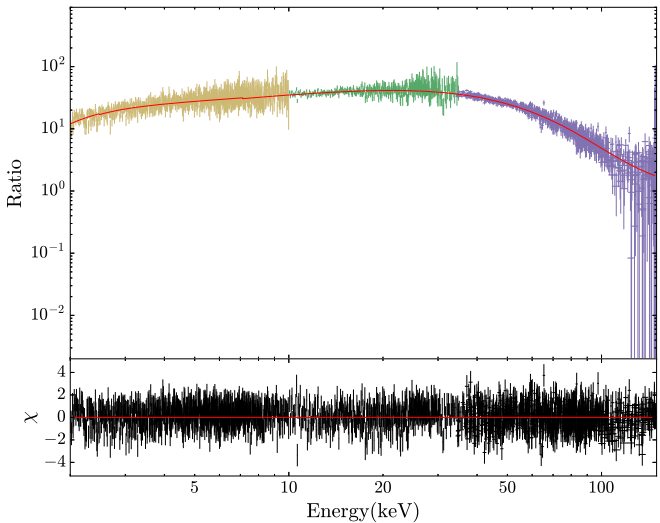

**Figure 4.** Ratio of the Swift J0243.6+6124 spectrum to that of the Crab pulsar in the 2–150 keV range obtained from HE (purple dotted line), ME (green dotted line), and LE (yellow dotted line) detectors of *Insight-HXMT* observation, represented by blue dots. The spectral ratio is reproduced with the model `TBabs*(cutoffpl+bbody)`. The reduced $\chi^2$(dof) is 0.86(2599).

The ratio of the pulsed to total flux can be used to estimate the fraction of the pulsed flux in several energy bands. We calculate it in three energy bands, i.e., 2–10 keV, 10–25 keV, and 25–150 keV. As shown in the bottom panel of Figure 3, the pulsed flux fraction in the entire *Insight-HXMT* energy band (2–150 keV) changed from ~30% to ~64% during the outburst. It is interesting to note that while the pulsed flux fraction in the soft band (2–10 keV) followed this trend and changed from ~8% to ~31%, the pulsed flux fraction in the hard band (10–150 keV), remained comparatively steady. At the end of the outburst, the pulsed flux fraction of the full energy band increased slightly.

### 2.3. Timing Analysis

Here we focus on spin evolution of the source throughout the outburst. First of all, for each observation, the spin period of the source is determined by using the epoch folding method. To reconstruct the intrinsic period of the neutron star, the orbital motion of the pulsar has to be, however, taken into account. Orbital parameters of the source have been reported by Doroshenko et al. (2018), Wilson-Hodge et al. (2018), and Ge et al. (2017). We use *Insight-HXMT* data to complement the available *Fermi*/GBM measurements[13] and improve orbital ephemerids by using the fitting process described in Weng et al. (2017) and Li et al. (2012). The resulting orbital solution is presented in Table 2. The spin period and its derivative in the pulsar's rest frame (the latter estimated from adjacent observations) are calculated using updated ephemerids and are shown in Figure 5.

We can see in Figure 5 that the pulsar exhibits strong spin-up throughout the outburst with the spin period decreasing from ~9.85 to ~9.79 s. The spin-up rate is correlated with flux, and rapidly reaches the maximum value $2.2(2) \times 10^{-8}$ s s$^{-1}$ close to the peak of the outburst. Then, it decreases steadily until it finally becomes comparable with zero, or even negative around MJD 58163 marked by a black arrow in Figure 5.

---
[13] http://gamma-ray.nsstc.nasa.gov/gbm/science/pulsars.html

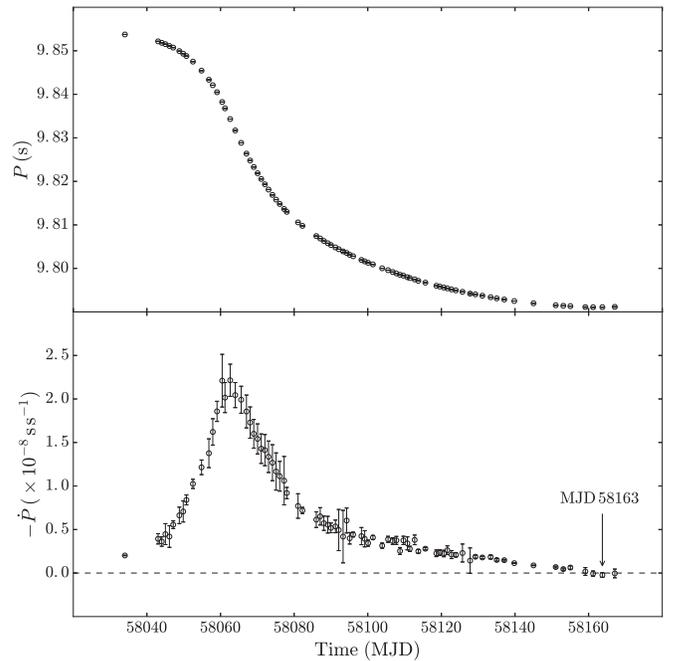

**Figure 5.** Intrinsic spin evolution (upper panel) and the derivative of the spin evolution (bottom panel) of Swift J0243.6+6124. The arrow denotes the time when $\dot{P} \approx 0$. The energy band is from 25 to 150 keV.

**Table 2**
The Position of the Source Was Determined By Kennea et al. (2017), and the Orbital Elements Were Calculated By Combining Data of *Fermi*/GBM Pulsar Project and *Insight-HXMT* Data

| Parameters | Value |
| --- | --- |
| R.A. | 02$^h$43$^m$40$^s$.33 |
| Decl. | 61°26′02″.8 |
| $P_{\rm orb}$, day | 27.8(6) |
| $a \sin i$, lt-s | 116.8(9) |
| $e$ | 0.09(5) |
| $\omega_0$, deg | −80(3) |
| $T_{\rm pa}$, MJD | 58,019.9(3) |

### 3. Application of the Accretion Torque Models

#### 3.1. Accretion Torque Models

The spin evolution of X-ray pulsars is driven by accretion torque and can be represented as (Ghosh et al. 1977),

$$-\dot{P} = \frac{NP^2}{2\pi I}, \qquad (1)$$

where $N$ and $I$ are the total torque and the effective moment of inertia of the neutron star respectively.

The torque can be written as (Ghosh & Lamb 1979),

$$N = n(\omega_s)\, \dot{M} \sqrt{GM_{\rm NS} r_{\rm m,d}}, \qquad (2)$$

where $\dot{M}$ is the mass accretion rate, $M_{\rm NS}$ is the mass of the neutron star, and $r_{\rm m,d}$ is the magnetospheric radius, which is considered to be the inner radius of the Keplerian disk. $n(\omega_s)$ is the dimensionless accretion torque, and it has a different form in different models. $\omega_s$ is the fastness parameter and is defined as the ratio of the neutron star's rotational velocity $\Omega_s$ to the Keplerian velocity $\Omega_K$ at $r_{m,d}$ (Ghosh & Lamb 1979).

There are several theories to estimate the magnetospheric radius $r_{m,d}$ in Equation (2). For example, in the model of Ghosh & Lamb (1979, hereafter the GL model), it can be determined from the Alfvén radius ($r_A$), through $r_{m,d} \simeq 0.52\, r_A$. $r_A$ is the radius where the ram pressure of the spherical freely infalling matter equals the magnetic pressure (Davidson & Ostriker 1972; Waters & van Kerkwijk 1989). In this model, the dimensionless torque $n(\omega_s)$ can be written as,

$$n(\omega_s) \approx 1.39 \times \frac{1 - \omega_s[4.03(1 - \omega_s)^{0.173} - 0.878]}{1 - \omega_s}. \quad (3)$$

However, some weaknesses of the GL model were pointed out (Wang 1987; Kluźniak & Rappaport 2007; Shi et al. 2015), e.g., the magnetic field is overestimated (Wang 1987, 1995).

In a more recent model by Shi et al. (2015, hereafter the SZL model), the improved magnetic field given by Wang (1995, 1996) is adopted. In this model three magnetospheric radii are considered ($r_{m1}$, $r_{m2}$, and $r_{m3}$). The dimensionless torques ($\omega_s \leqslant 1$) in this model can be written as,

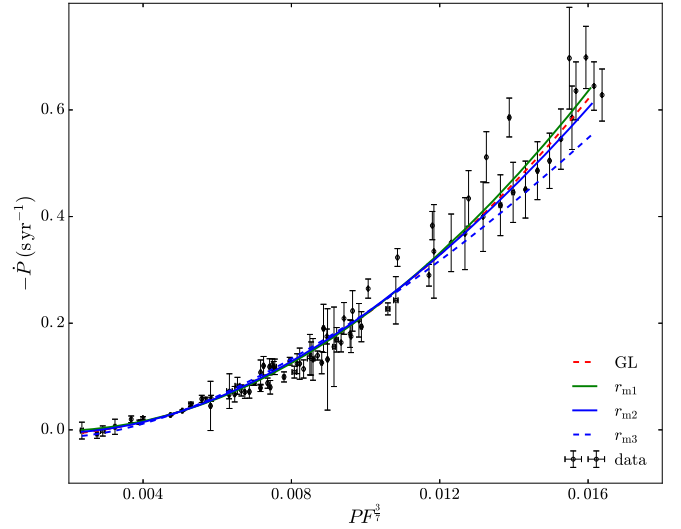

**Figure 6.** Relation between the spin-up rate $-\dot{P}$ and $PF^{3/7}$ during the giant outburst. Open black circles denote the data of *Insight*-HXMT. The red dashed line denotes the fitting result of the GL model. The green-solid line, the blue-solid line and the blue-dashed line denotes the fitting lines of $r_{m1}$, $r_{m2}$, and $r_{m3}$ in the SZL model, respectively.

$$n(\omega_s) = \begin{cases} r_{m1}: & (1 - \omega_s) + \frac{\sqrt{2}}{3}(\frac{2}{3} - 2\omega_s + \omega_s^2), \\ r_{m2}: & (1 - \omega_s) + 314.258 * f^{34/10} P_1^{-1/12} L_{37}^{-3/20} \omega_s^{-1/12}(\frac{2}{3} - 2\omega_s + \omega_s^2), \\ r_{m3}: & (1 - \omega_s) + 543.248 * P_1 \omega_s(\frac{2}{3} - 2\omega_s + \omega_s^2), \end{cases} \quad (4)$$

where $f = \left(1 - \sqrt{\frac{R}{r}}\right)^{1/4}$. $r_{m1}$ is the Alfvén radius ($r_A$), corresponding to the magnetospheric radius $r_{m,d}$ in the GL model (at variance with the following cases, this one is referred to as the radius of an uncompressed magnetic field). $r_{m2}$ is the magnetospheric radius when the compression of the outer magnetosphere (outside $r_{m2}$) by accreting matter is taken into account (Shi et al. 2014, 2015). $r_{m3}$ is the magnetospheric radius when the compression of the entire magnetosphere by accreting matter is taken into account (Kulkarni & Romanova 2013). $P_1$ and $L_{37}$ are the spin period in units of s and the luminosity in $10^{37}$ erg s$^{-1}$, respectively.

The X-ray luminosity due to the accretion of matter on the neutron star in the above models can be derived (Ghosh & Lamb 1979; Shi et al. 2015) via,

$$L = GM_{NS}\dot{M}/R, \quad (5)$$

where $R$ is the radius of the neutron star. Assuming the observed flux $F$ reflects the luminosity, then $F = L/(4\pi D^2)$, where $D$ is the distance to the source. The characteristic values of neutron star that $M_{NS} = 1.4\, M_\odot$ and $R = 10^6$ cm were applied, where $M_\odot$ is the mass of the Sun. After that, the above Equations can then be used to fit the dependence of the spin-up rate on flux, and estimate the distance and the magnetic field strength of the neutron star.

### 3.2. Fitting Results

As discussed above, the relation of spin-up rate and flux is shown in Figure 6. All models adequately describe the spin-up at low accretion rates, and the differences only appear at high rates.

Fitting results of different models are shown in Table 3. For the two uncompressed magnetic field models, i.e., the GL model and $r_{m1}$ in the SZL model, the distance has a similar value and agrees with the lower limit of 5.0 kpc at $\geqslant$99% confidence level given by van den Eijnden et al. (2018). The magnetic field strength in the latter model is higher than the former one by a factor of 2, and they both are in line with conclusions by Doroshenko et al. (2018) and Tsygankov et al. (2018), who suggested $\sim 10^{13}$ G. On the other hand, results of compressed models of $r_{m2}$ and $r_{m3}$ in the SZL model show much shorter distance than the uncompressed models, which appear to be at odds with the *Gaia* distance estimate. The magnetic field strength of $r_{m2}$ is close to the uncompressed models, but for $r_{m3}$, it is much weaker.

### 4. Discussion and Summary

The fact that the pulsed flux fraction in broad and soft energy bands in Figure 3 follows the same trend suggests that most of the pulsed flux ($\sim$50%) actually comes from the soft band. On the other hand, other patterns of evolution for the hard band

Table 3
Fitting Results of Two Accretion Torque Models

| Model | $B$ (G) | $D$ (kpc) | $L_{max}$ (erg s$^{-1}$) | $\chi^2$ |
|---|---|---|---|---|
| GL | $(5.98 \pm 0.20) \times 10^{12}$ | $6.81 \pm 0.04$ | $(1.83 \pm 0.02) \times 10^{39}$ | 1.5 |
| $r_{m1}$ | $(1.02 \pm 0.04) \times 10^{13}$ | $5.08 \pm 0.04$ | $(1.02 \pm 0.01) \times 10^{39}$ | 1.4 |
| $r_{m2}$ | $(1.49 \pm 0.04) \times 10^{13}$ | $0.42 \pm 0.01$ | $(6.96 \pm 0.14) \times 10^{38}$ | 1.6 |
| $r_{m3}$ | $(1.07 \pm 0.02) \times 10^{11}$ | $0.032 \pm 0.001$ | $(4.06 \pm 0.06) \times 10^{34}$ | 2.3 |

**Note.** $\chi^2$ denotes the reduced $\chi^2$ of the fitting.

might suggest that the emission mechanisms are different between these two energy bands at the epoch of the peak. Such change is likely associated with the change of the emission region geometry, i.e., onset and growth of the accretion column. Similar conclusions were made by Doroshenko et al. (2018) based on the comparison between the pulsed *Fermi*/GBM and unpulsed *Swift*/BAT fluxes. Furthermore, they found that the pulse profile at high fluxes is double-peaked. At the same time, the bottom panel of Figure 3 shows that the pulsed flux fraction in 2–150 keV reaches the peak. It is also interesting to note that van den Eijnden et al. (2018) found a radio jet after the epoch of the peak of this outburst, i.e., the formation of the jet coincides with softening of the X-ray spectrum. While the jet must be formed far away from the neutron star (van den Eijnden et al. 2018), it might still be possible that the two phenomena might be related. For instance, Illarionov & Kompaneets (1990) suggested that heating of the accretion flow by X-rays from the pulsar might lead to the formation of outflows, which is more likely in the case of super-critical accretion and might also play a role in the jet formation or collimation. In this source, the luminosity is far more than the critical X-ray luminosity in Illarionov & Kompaneets (1990). However, the source persists spinning-up until the end of the outburst. The reason is that although the "heated wind" contributes to the drop of spin-up rate as shown in the bottom panel of Figure 5, the total accretion torque is larger and accelerates the neutron star.

In van den Eijnden et al. (2018), the source reached the super-Eddington regime ($2 \times 10^{38}$ erg s$^{-1}$) during the outburst. However, it is reasonable to apply the models mentioned above even though such a high luminosity is not considered by the models because for the majority of the time ($\sim$85%) the observed flux is below the Eddington limit. In addition, the strong magnetic field causes the effective electron scattering cross-section perpendicular to the field lines to become lower, and the photons can effectively escape from the walls of the accretion column (Basko & Sunyaev 1976; Lyubarskii & Syunyaev 1988; Mushtukov et al. 2015). Then, in this strong magnetic field regime, Equation (5) can provide an approximate expression of the correlation between the luminosity and accretion rate. But, close to the peak of the outburst, the results presented here should be considered as approximate. The discrepant result between compressed models and uncompressed models, which are consistent with *Gaia* data might point to the fact that the field of the source is indeed stronger than for most BeXRBs so that the magnetosphere is not significantly compressed and thus compressed magnetosphere torque models are not applicable in this case.

In summary, we presented our analysis of *Insight*-HXMT data on the Be/X-ray pulsar Swift J0243.6+6124 during the 2017–2018 outburst. The broadband spectra (2–150 keV) of the source can be described with a cutoff power-law continuum with an additional soft blackbody component and a gaussian profile. We found that variations of the pulsed flux fraction with time are different in the three energy bands, which are likely related to changing patterns of the pulse profile reported in Tsygankov et al. (2018) and associated with the onset of accretion column. No evidence is found for cyclotron line in the spectra of *Insight*-HXMT; perhaps there is no cyclotron resonant scattering process during this outburst, or it occurs at an energy higher than the maximum energy range of *Insight*-HXMT. We estimated the magnetic field with two accretion torque models (GL and SZL models). The results confirm that this source is a ULX pulsar with $B \sim 10^{13}$ G and $L > 10^{39}$ erg s$^{-1}$ ($D > 5$ kpc).

We thank the anonymous referee and Dr. Qingcui Bu for useful comments. This work made use of the data from the *Insight*-HXMT mission, a project funded by the China National Space Administration (CNSA) and the Chinese Academy of Sciences (CAS). The *Insight*-HXMT team gratefully acknowledges the support from the National Program on Key Research and Development Project (grant No. 2016YFA0400800) from the Minister of Science and Technology of China (MOST) and the Strategic Priority Research Program of the Chinese Academy of Sciences (grant No. XDB23040400). The authors are thankful for support from the National Natural Science Foundation of China under grant Nos. 11503027, 11673023, 11733009, U1838108, U1838201, and U1838202; and Russian Science Foundation grant 19-12-00423.

ORCID iDs

Yue Zhang https://orcid.org/0000-0003-0396-2689
Lian Tao https://orcid.org/0000-0002-2705-4338
Long Ji https://orcid.org/0000-0001-9599-7285
ChangSheng Shi https://orcid.org/0000-0002-9811-537X
LinLi Yan https://orcid.org/0000-0002-2244-4222